\documentclass[aps,prb,twocolumn,superscriptaddress,nobibnotes]{revtex4-2}

\usepackage{graphicx}
\usepackage{physics}
\usepackage{amssymb}
\usepackage{amsmath}
\usepackage{color}
\usepackage{dcolumn}
\usepackage{epstopdf}
\usepackage{bm}
\usepackage{natbib}
\usepackage{tikz}
\usepackage[colorlinks=true,citecolor=blue]{hyperref}

\begin{document}

\title{Heat Conduction and Energy Relaxation in an InAs Nanowire Approaching the Clean One-Dimensional Limit}

\author{Subhomoy Haldar}
\email{shaldar@iitk.ac.in}
\affiliation{NanoLund and Solid State Physics, Lund University, Box 118, 22100 Lund, Sweden}
\affiliation{Department of Physics, Indian Institute of Technology Kanpur, Uttar Pradesh 208016, India }
\author{Diego Subero}
\affiliation{Pico group, QTF Centre of Excellence, Department of Applied Physics, Aalto University School of Science, P.O. Box 13500, 00076 Aalto, Finland
}
\author{Mukesh Kumar}
\affiliation{NanoLund and Solid State Physics, Lund University, Box 118, 22100 Lund, Sweden}
\author{Bayan Karimi}
\affiliation{Pico group, QTF Centre of Excellence, Department of Applied Physics, Aalto University School of Science, P.O. Box 13500, 00076 Aalto, Finland
}
\affiliation{Pritzker School of Molecular Engineering, University of Chicago, Chicago IL 60637, USA}
\author{Adam Burke}
\affiliation{NanoLund and Solid State Physics, Lund University, Box 118, 22100 Lund, Sweden}

\author{Lars Samuelson}
\affiliation{NanoLund and Solid State Physics, Lund University, Box 118, 22100 Lund, Sweden}
\affiliation{Institute of Nanoscience and Applications, SUSTech, Shenzhen, China}
\author{Jukka Pekola}
\affiliation{Pico group, QTF Centre of Excellence, Department of Applied Physics, Aalto University School of Science, P.O. Box 13500, 00076 Aalto, Finland
}
\author{Ville F. Maisi}
\email{ville.maisi@ftf.lth.se}
\affiliation{NanoLund and Solid State Physics, Lund University, Box 118, 22100 Lund, Sweden}

\date{\today}

\begin{abstract}

We investigate heat conduction and energy relaxation in an InAs semiconductor nanowire using a hybrid semiconductor–superconductor architecture. Local electronic temperatures are measured with an in-situ grown quantum dot thermometer, while controlled Joule heating is applied at different locations along the wire to probe temperature gradients at sub-kelvin temperatures. With a one-dimensional heat transport model, we calculate an electron-phonon heat flow that scales as $Q_{\mathrm{e-ph}} \propto T^{2.6}$, which is in close agreement with the $T^3$ dependence predicted for a clean one-dimensional electron gas coupled to a phonon bath. We further estimate a characteristic length $l_{\rm eq} = 370~\mathrm{nm}$, beyond this length scale, phonon-mediated heat transport dominates over heat conduction in our nanowire. Our results provide a quantitative measure of energy relaxation mechanisms in a one-dimensional semiconductor and provide a framework for studying heat flow in low-dimensional nanostructures.

\end{abstract}

\maketitle

\section{Introduction}

Heat conduction and energy dissipation in nanoscale systems play a central role in the performance of electronic, thermoelectric, and quantum devices~\cite{mosso2017, pekola2015, chen2018, shi2020, Kumar2024}. At the heart of these devices, energy transport is determined by the electrical and thermal conductivity, phonon-assisted processes, and energy dissipation to the environment~\cite{Giazotto2006}. As the size of the system approaches the electronic mean free path and electron-phonon scattering length, energy transport depends on the quantized energy modes and mesoscopic effects~\cite{Battisti2024, chen2021, Majidi2022}. A quantitative understanding of these regimes is essential for nanoscale thermoelectric devices, management of heat dissipation in cryogenic electronics, and mitigating decoherence in quantum circuits~\cite{Mickelsen2023, Undseth2023, Prete2019, pekola2015}. In particular, semiconductor nanowires (NWs) provide a platform for exploring energy transport at these length scales. Their one-dimensional geometry, high crystalline quality, and tunability of electronic-transport by electrostatic gates help control charge and heat conduction independently~\cite{Kumar2024, Majidi2022, Prete2019}. In lower dimensions, phonon spectra are modified, boundary scattering becomes dominant, and cooling power laws can deviate substantially from the conventional $T^5$ dependence for ordinary metals~\cite{Hofmann2020, Matthews2012, Wang2025}. Such deviations, together with violations of the Wiedemann–Franz law in nanoscale semiconductors and metals due to Coulomb effects~\cite{Majidi2022, Kubala2008} show a different thermal transport regime. Recent years have seen rapid progress in the study of heat conduction and phonon-mediated relaxation in low-dimensional semiconductors and van der Waals materials~\cite{karimi2021, Wang2025, Majidi2022}. Control of mesoscopic heat flow has enabled the realization of quantum-dot (QD) heat engines operating near thermodynamic efficiency limits~\cite{josefsson2018}, and advances in cryogenic calorimetry, bolometry, and noise thermometry~\cite{Chawner2021, karimi2021, van2024}. Nevertheless, quantitative measurements of heat transport and electron–phonon interactions in semiconductor NWs remain challenging. A key experimental difficulty lies in the ability to inject well-defined heat into a nanoscale device while simultaneously performing local thermometry under thermal isolation.

In this work, we investigate the heat conduction and energy dissipation processes by using a hybrid semiconductor–superconductor architecture in which an epitaxial InAs NW hosts in-situ grown QDs that work as a primary electronic thermometer. Superconducting contacts with negligible thermal conductance provide localized Joule heating while maintaining electrical control. By measuring the source and drain temperatures of the QD under steady-state heating, we map the spatial temperature profile along the wire. The resulting data are analyzed using a one-dimensional heat-transport model that incorporates both electronic diffusion and generalized $P\propto T^{n}$ electron–phonon cooling. This allows us to calculate key material and transport parameters, including the electron–phonon coupling strength, the dimensionality-dependent power-law exponent, and the characteristic thermal relaxation length.

\begin{figure*}
\centering
\includegraphics[width=1.05\textwidth]{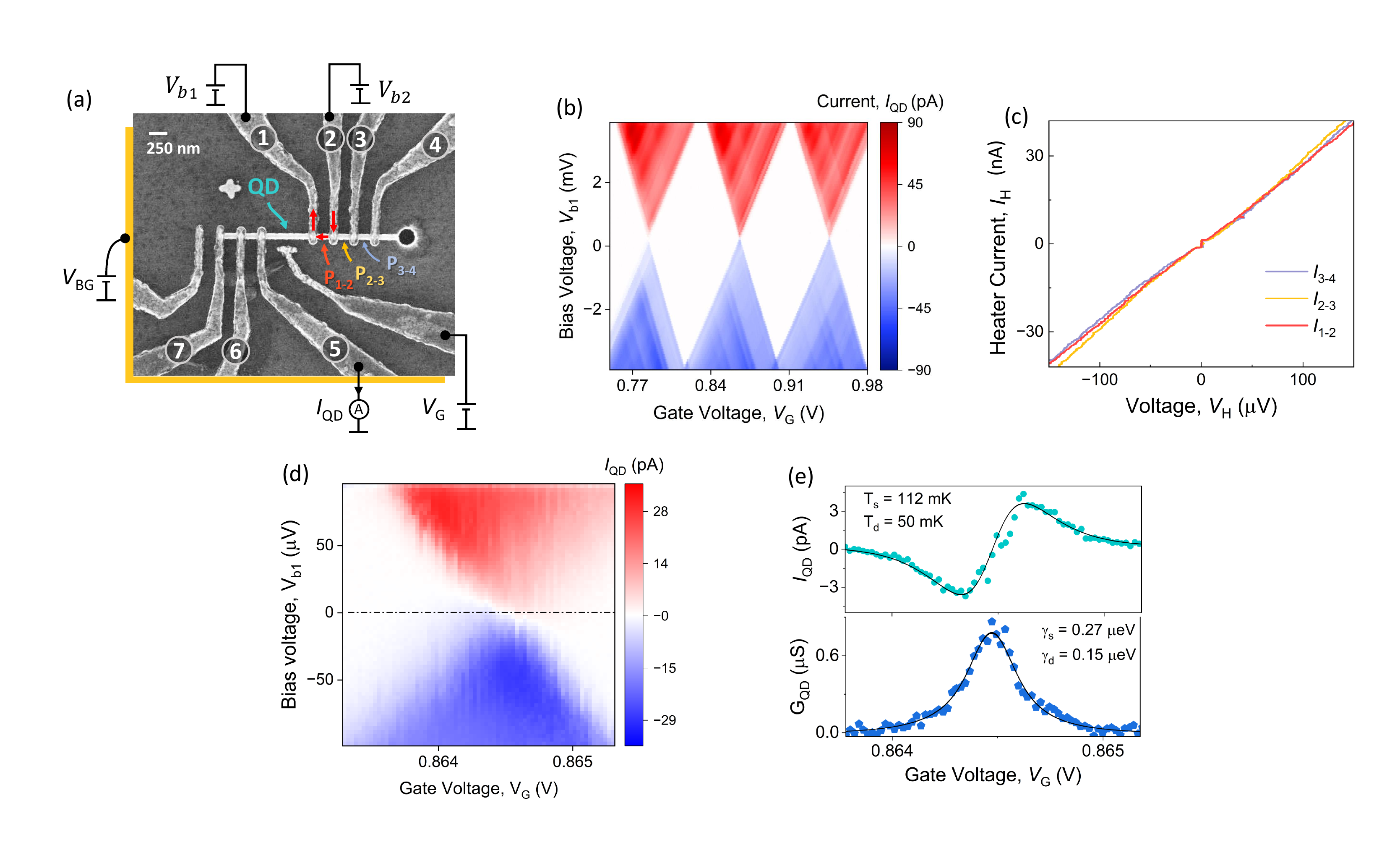}
\vspace{-1.5cm}
\caption{(a) Scanning electron micrograph of the semiconductor-superconductor hybrid device showing the NW-QD with the superconducting contacts and measurement setup. The QD is located 1.3~$\mu$m away from the tip. The contacts are positioned 250 nm apart, which allows for localized Joule heating by passing current between different pairs of contacts. (b) The measured current through the QD as a function of bias voltage $V_{b1}$ and side-gate voltage $V_G$, showing the standard Coulomb diamonds, and (c) Current-bias voltage characteristics of the S-Sm-S contacts. (d) A zoomed-in image of the Coulomb peak, measured with 25~fW heating power applied to the 1-2 lead, and (e) the current and differential conductance data as a function of gate voltages at $V_{b1} = 0.$ The solid lines are fits to Eq.~(\ref{Land}). }\label{fig1}
\end{figure*}

\section{Experimental Details}

Our hybrid device comprises a wurtzite InAs semiconductor NW, 70~nm in diameter, which is grown using the chemical beam epitaxy (CBE) method~\cite{Fasth2007}. During the NW growth, two $\sim$10~nm long InP barriers were grown to confine electrons within a $\sim$100~nm long InAs segment, forming a QD. The QD is located approximately 1.3~$\mu$m from the tip. As shown in Fig.~\ref{fig1}(a), seven galvanic contacts, separated by $\sim$250~nm, are made on the NW using electron beam lithography (EBL) followed by thermal evaporation of 1~nm Ti/120~nm Al. These superconducting contacts are used to heat specific segments of the NW and measure electronic transport through the QD. Because at mK temperatures the heat conductance of superconductors is very low, the electrons in the semiconductor are thermally insulated, such that the heat can only flow and be dissipated within the NW. Notably, during NW growth, an Au particle is used as a seed, which remains at the tip of the wire and forms a galvanic contact with the wire. To eliminate any unwanted heat dissipation by the Au particle, we remove it using an additional round of EBL followed by argon ion milling; the resulting dark spot at the tip of the NW is visible in Fig.~\ref{fig1}(a). The device is fabricated on a degenerately doped Si substrate with a 200~nm thick thermally grown SiO$_2$. The conducting substrate acts as a global back gate ($V_\text{BG}$), which helps in tuning the electrical conductivity of the entire wire and the QD. In addition, the side gate voltage ($V_\text{G}$) controls the electrochemical potential ($\varepsilon_\text{N}$) of the QD directly. For the presented experiments, we use $V_\text{BG}$ = 4.5~V and vary $V_\text{G}$. Measurements are performed in a dilution refrigerator with electronic temperature, $ T_e\approx$~40 mK.

Similar NWs have been used previously for the thermoelectric devices and heat-transport measurements, both in the ballistic and diffusive regimes~\cite{Svilans2018, Wu2013, Kumar2024}. Transport experiments show that these wires have an electronic mean free path, $l_f\sim$100 nm~\cite{Hansen2005, Kumar2024}.

\section{Results and Discussion}

We start by measuring the electronic transport properties of the QD in the absence of external heating. A bias voltage $V_{\mathrm{b1}}$ is applied to lead 1 and the resulting current $I_{\mathrm{QD}}$ is measured through lead 5, as illustrated in Fig.~\ref{fig1}(a). All other contacts are left electrically floating to avoid spurious Joule heating. Figure~\ref{fig1}(b) shows the measured current as a function of $V_{\mathrm{b1}}$ and side-gate voltage $V_G$, showing well-defined Coulomb diamonds characteristic of a QD. The gate voltage linearly tunes the electrochemical potential of the dot, $\varepsilon_N = \alpha_G V_G$, with a lever arm $\alpha_G = 49~\mu\mathrm{eV/mV}$. From the periodicity of the Coulomb peaks, we determine the charging energy $E_c = 4~\mathrm{meV}$. The nearly uniform peak separation indicates that the orbital level spacing is significantly smaller than $E_c$, consistent with the size of the dot in our experiment, $l_\text{QD}\sim$100~nm .

The current–voltage characteristics of the superconductor-NW-Superconductor segments used as heaters are shown in Fig.~\ref{fig1}(c). All segments exhibit weak proximity-induced superconductivity with a critical current $I_c \approx 1.2~\mathrm{nA}$, followed by an approximately linear resistive regime at higher bias. The reproducible and nearly identical behavior allows the contacts to be used as local Joule heaters with known input power.

\subsection{Quantum-dot thermometry}

To probe local electronic temperatures, we utilize the QD as a primary thermometer. When a heating voltage $V_H$ is applied between a selected pair of contacts, the resulting temperature rise at the source and drain barriers of the QD, $T_s$ and $T_d$ respectively, is calculated from the thermally broadened Coulomb peak. The QD operated in the thermally broadened regime has previously been used for thermometry~\cite{hoffmann2009, ahmed2018}. In our configuration, the bias voltage applied to the QD is set by $V_{\mathrm{b1}}$, while the heating power is determined by $V_H = V_{\mathrm{b2}} - V_{\mathrm{b1}}$ [Fig.~\ref{fig1}(a)]. Figure~\ref{fig1}(d) shows a representative charge-stability map obtained with a constant heating power of $25~\mathrm{fW}$ applied between leads 1 and 2. The corresponding zero-bias current and conductance are shown in Fig.~\ref{fig1}(e). To extract $T_s$ and $T_d$, we fit the measured $I_{\mathrm{QD}}(V_G)$ and $G_{\mathrm{QD}}(V_G)$ using the Landauer–B\"uttiker formalism for a single-level QD,

\begin{eqnarray}\label{Land}
I_{\mathrm{QD}} &=& \frac{e}{h}\int \tau(E)\left[f_s(E)-f_d(E)\right] dE, \nonumber\\
G_{\mathrm{QD}} &=& \frac{e^2}{h}\int \tau(E)\left[\frac{df_s}{dE}-\frac{df_d}{dE}\right] dE,
\end{eqnarray}

where $\tau(E)=2\Gamma\gamma/[\{E+\alpha_G e(V_G^0-V_G)\}^2+\gamma^2]$ is the energy-dependent transmission function with $\gamma = (\gamma_s+\gamma_d)/2$, $\Gamma=\gamma_s\gamma_d/(\gamma_s+\gamma_d)$ and $f_{s(d)}$ are Fermi functions at temperatures $T_{s(d)}$. The black solid curves in Fig.~\ref{fig1}(e) show the representative fitted results using Eq.~(\ref{Land}). We here calculate the following four parameters: temperatures $T_s$ and $T_d$, and tunnel couplings $\gamma_s$ and $\gamma_d$ by fitting: (i) photocurrent maxima and their spread, and (ii) conductance peak-height and width. The calculated parameter values are shown in the inset. We note that the energy of electrons at the base temperature $k_BT_e\sim$ 4~$\mu$eV is an order of magnitude higher than the energy scale set by the tunnel couplings $\gamma_s$ and $\gamma_d$ ($\sim$ 0.3 $\mu$eV). The device is intentionally tuned to the regime $\gamma_{s,d} \ll k_B T$, ensuring that the Coulomb peak shape is dominated by thermal broadening and thus sensitive to temperature.

\begin{figure}
\centering
\includegraphics[width=0.45\textwidth]{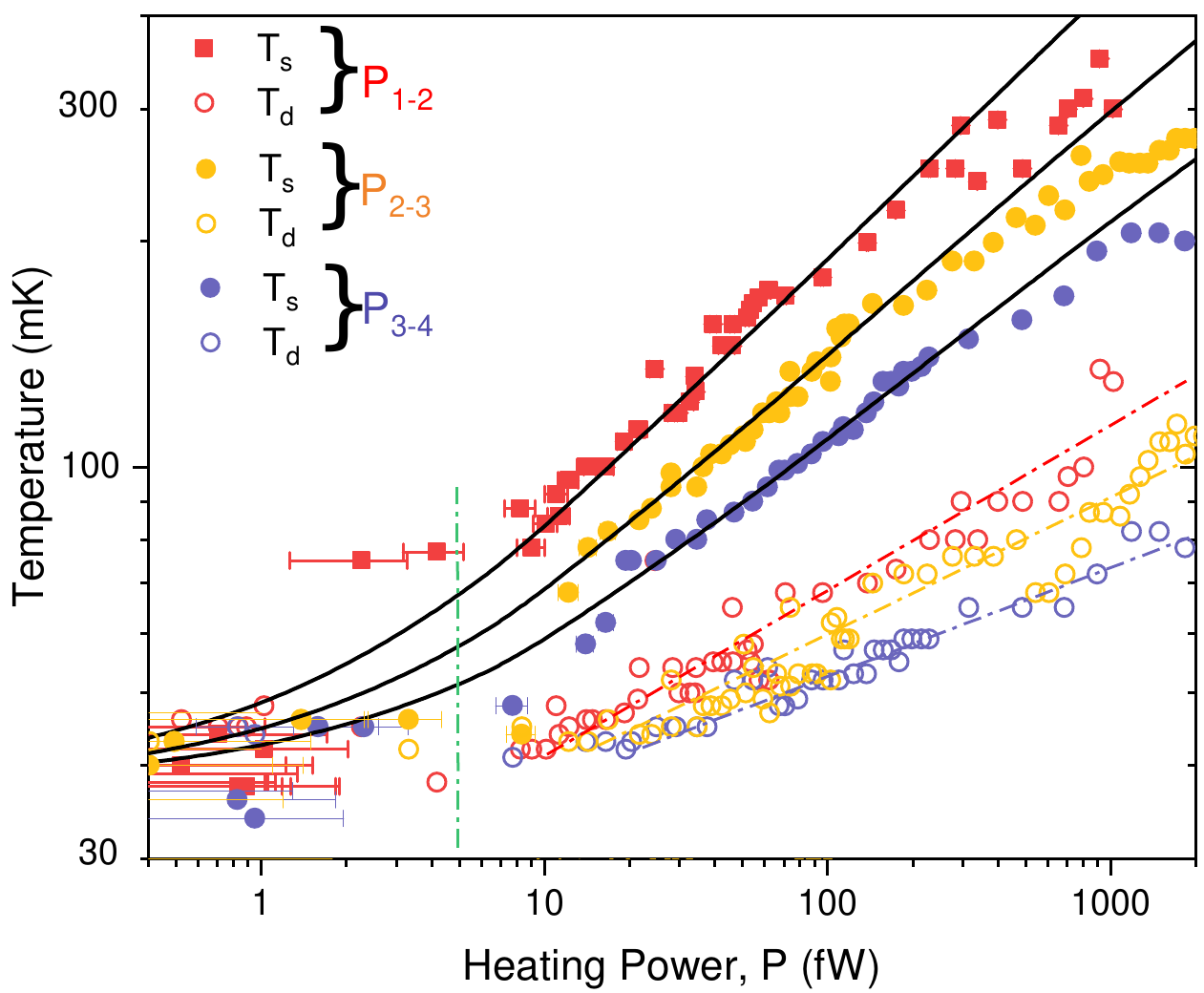}
\caption{Measured source ($T_s$) and drain ($T_d$) temperatures as a function of the heating power applied to the nanowire segments between leads 1–2, 2–3, and 3–4. The source electrode is on the heated side. Solid lines represent fits obtained using the one-dimensional heat-flow model of Eq.~(\ref{heatCond}) with $\Sigma = 2\times10^9$~W/m$^3$K$^{2.6}$. The dashed curves are added as guides to the eye. The green dashed line marks the transition from the supercurrent regime to the Joule-heating regime in the S–Sm–S heaters.}\label{fig2}
\end{figure}

Using this thermometry technique, we measure $T_s$ and $T_d$ as functions of the heating power applied to three different NW segments. The results are summarized in Fig.~\ref{fig2}. When the heaters operate in the superconducting regime, no measurable temperature rise is observed, and both $T_s$ and $T_d$ remain at the base temperature of approximately $38~\mathrm{mK}$. Upon switching to the resistive regime, heating powers of $\sim10~\mathrm{fW}$ result in measurable temperature increase. We note that, for a given input power, the measured $T_s$ is highest when heat is injected closest to the QD and decreases as the heating location is moved farther away. This demonstrates the presence of longitudinal temperature gradients along the NW even under steady-state conditions. At higher powers, a small but measurable increase in $T_d$ is also observed, indicating a weak heat leak through the QD during thermometry. This effect is analyzed quantitatively below.

\subsection{One-dimensional heat-transport model}

To analyze the measured temperature profiles, we use a one-dimensional heat-flow model. For our NW with a diameter of $\sim$70~nm, the electron and phonon modes are quantized in the transverse direction; therefore, a 1D heat transport model can be considered as a good approximation~\cite{Hansen2005, chen2018}. We here discretize the NW into elementary lengths, \textit{dx}, with $T(x)$ being the temperature at position \textit{x}, see Fig.~\ref{fig3}. The heat balance equation with the input heat ${dP}$ to an elementary volume $dv=A \cdot dx$ can be expressed as:

\begin{equation}\label{totalBalance}
    dP = dQ_\text{e-ph} + dQ_c.
\end{equation}

Here, $dQ_\text{e-ph}$ is the heat dissipation due to the electron-phonon interaction and $dQ_c$ is the heat conduction along the NW. The first term of Eq.~(\ref{totalBalance}) can be written as:

\begin{figure}
\centering
\includegraphics[width=0.47\textwidth]{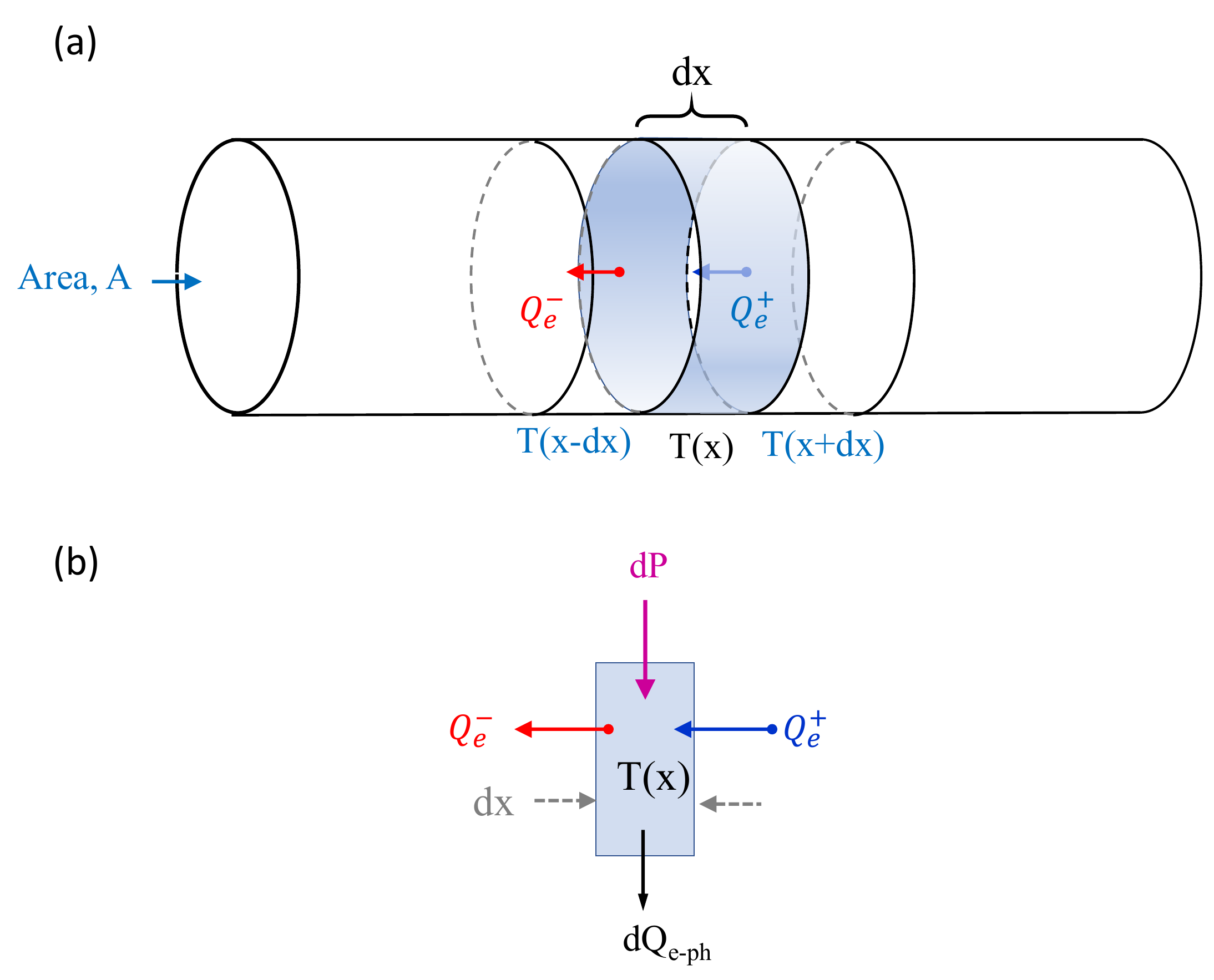}
\caption{(a) Schematic of a one-dimensional NW with cross-sectional area \textit{A}, which consists of an elementary length \textit{dx}, and the temperature at position \textit{x} is denoted by \textit{T(x)}. (b) The heat balance for an elementary volume showing the heating power injected as the sum of heat dissipation via electron-phonon interaction and heat conduction to the neighboring elements.}\label{fig3}
\end{figure}

\begin{equation}\label{electronPhonon}
    dQ_\text{e-ph} = \Sigma ~ dv \left[T(x)^n - T_p^n\right].
\end{equation}

where $\Sigma$ is the electron-phonon coupling constant and $T_p$ is the phonon temperature. 

On the other hand, the net heat flow through an elementary volume is $dQ_c = (Q_e^- - Q_e^+)$, where $Q_e^+$ and $Q_e^-$ are the heat entering and leaving the volume, respectively, as illustrated in Fig.~\ref{fig3}(a) and (b). By assuming Wiedemann-Franz like electrical heat conduction, this is given by,

\begin{eqnarray}\label{heatFlow}
    Q_e^{\pm} = \pm\frac{L \sigma}{2} \left[\frac{ T(x\pm dx)^2 - T(x)^2}{dx}\right] A .
\end{eqnarray}

Here, the thermal conductivity, $\kappa_e(x) = L \sigma T(x)$, with $\sigma$ being the electrical conductivity and $L =\xi L_0$, with the Lorentz number $L_0 = 2.44\times 10^{-8}$~W$\Omega$/K$^2$ and $\xi$ is the effective Lorenz factor renormalization in our 1D diffusive heat transport model that has the value $\xi= 1$ for the Wiedemann-Franz law.
 
Taylor expanding each term of Eq.~(\ref{heatFlow}) around $x$  and neglecting higher-order terms, we have

\begin{equation}\label{heatConductionTaylor}
    dQ_c = -\left[\left\{\frac{dT(x)}{dx}\right\}^2dx^2+T(x)\frac{d^2T(x)}{dx^2} dx^2\right]\frac{L\sigma A}{dx}.
\end{equation}

Finally, using Eqs.~(\ref{totalBalance}), (\ref{electronPhonon}) and (\ref{heatConductionTaylor}), we can express the heat balance equation:

\begin{eqnarray}\label{heatCond}
    \frac{p}{L \sigma} = -\left[ \left\{\frac{dT(x)}{dx}\right\}^2 +  T(x) \frac{d^2T(x)}{dx^2} \right] \\ \nonumber
    + \frac{\Sigma}{L\sigma} \bigg[T(x)^n - T_p^n\bigg].
\end{eqnarray}

Here, $p = {dP}/{dv}$ is the input heat power density.\\

\subsection{Nanowire thermal properties}

We fit the measured $T_s(P)$ in Fig.~\ref{fig2} using Eq.~(\ref{heatCond}) with $\Sigma$, \textit{n}, and the Lorentz factor $\xi$ as the fitting parameters. We here use the electrical conductivity of the NW, $\sigma$= $1.85\times 10^4$~$\Omega^{-1} \text{m}^{-1}$, calculated from the linear response regime of Fig.~\ref{fig1}(c). The black curves in Fig.~\ref{fig2} show the self-consistently fitted results with $n = 2.6\pm 0.2$, $\Sigma = (2\pm 0.2)\times 10^9$ W/m$^3$K$^{2.6}$ and $\xi=0.4\pm0.1$. A 60\% reduction in the Lorenz factor shows a factor of two deviation from the Wiedemann–Franz law. The proximity effect inducting partial superconductivity from the leads to the NW is a likely reason for this suppressed heat conductance~\cite{Peltonen2010}. 

The value of $n$ provides insight into the underlying electron--phonon coupling mechanism. In a clean low-dimensional conductor, the heat flow follows $Q_{\mathrm{e\text{-}ph}} \propto T^{d+2}$, where $d$ is the effective dimensionality~\cite{stepanov2023}. For a one-dimensional nanowire ($d=1$), the clean-limit prediction is $Q_{\mathrm{e\text{-}ph}} \propto T^{3}$, close to our experimentally obtained exponent ($n = 2.6$). By contrast, in the disorder-dominated regime, $q l_{ph} \ll 1$, theory predicts $n \approx 4$~\cite{altshuler1985,Rammer1986,Sergeev2000}, where $q \sim k_B T/\hbar c_s$ is the phonon wave vector and $l_{ph}$ is the phonon mean free path. Using $c_s \approx 4200\,\mathrm{m/s}$~\cite{Battisti2024} and $l_{ph} \approx 250\,\mathrm{nm}$~\cite{Dorsch2021, Swinkels2015} for InAs NW, we estimate $q l_{ph} \approx 1.5$ at $T = 100\,\mathrm{mK}$. This places our device outside the disorder-dominated transport and supports clean one-dimensional electron-phonon coupling; consistent with the criterion $ql_{ph}\geq 1$~\cite{altshuler1985,Rammer1986,Sergeev2000}. Furthermore, the slight reduction of the exponent below the ideal $T^{3}$ dependence may originate from the surface or interface scattering, which is known to renormalize the temperature scaling toward lower values~\cite{Sergeev2000, Ceder2005}.

Next, to realize the relative impact of electronic heat conduction and electron-phonon energy relaxation, we estimate the corresponding thermal conductance, $G_{c}^t$ and $G_{\rm e-ph}^t$. Using the Wiedemann-Franz relation, we calculate the thermal conductivity, $\kappa_e = \xi L_0\sigma T = 1.8\times10^{-5}\,\mathrm{W\,m^{-1}K^{-1}}$ at $T = 100~\mathrm{mK}$, and for a representative segment of length $l = 250~\mathrm{nm}$, the electronic thermal conductance $G_{c}^t = \kappa_e A/l = 2.8\times10^{-13}\,\mathrm{W\,K^{-1}}$. In comparison, the thermal conductance due to electron-phonon coupling in the same segment is $G_{\rm e-ph}^t = n\Sigma V T^{n-1} = 1.3\times10^{-13}\,\mathrm{W\,K^{-1}}$ at $100~\mathrm{mK}$. The estimated $G_{c}^t$ is more than a factor of 2 higher than $G_{\rm e-ph}^t$ at this length scale, which implies that heat is removed more efficiently along the electronic channel than it is transferred to the phonon bath.

The above estimates also allow us to calculate the characteristic length scale at which electronic heat conduction and electron-phonon energy relaxation contribute equally to thermal transport. As the electronic thermal conductance scales as $G_{c} \propto l^{-1}$, whereas the electron-phonon thermal conductance scales as $G_{\rm e\text{-}ph} \propto l$, equating the two yields

\begin{equation}
    \ell_{\rm eq} = \sqrt{\frac{\xi \sigma L_{0}}{n\Sigma\,T^{\,n-2}}},
\end{equation}

which for our NW is found to be $\ell_{\rm eq} \approx 370~\mathrm{nm}$ at 100~mK. This sets the length scale over which both the thermal conductances compete. For segments shorter than $\ell_{\rm eq}$, electronic conduction dominates, whereas for longer segments electron–phonon relaxation determines the thermal response. The length of our NW is larger than $\ell_{\rm eq}$, which explains the spatial temperature variations observed experimentally.


\subsection{Quantum Dot Thermometer}

We next turn to considering the QD thermometer and estimating the heat leak, $P_L$, through the dot during temperature measurements. We here use Eq.~(\ref{heatCond}) to determine the value of $P_L$ required to reproduce the measured drain-side temperature $T_d$ shown in Fig.~\ref{fig2}. We use the same parameter values $\Sigma$, $n$, and $\xi$ that were determined for the source side. Figure~\ref{fig4} presents the calculated $P_L$ as a function of the input heat. The black line in Fig.~\ref{fig2}(a) show the $P_L = k~P$ variation with $k=0.009$. This confirms that the QD thermometer contributes less than 1~\% parasitic heat flow, demonstrating it as a minimally invasive probe for local electronic temperature measurement.

Using the measured source and drain temperatures across the 100~nm long QD and the estimated heat leak $P_L$, we calculate the thermal conductance of the dot. For $T_s = 200~\mathrm{mK}$, $T_d = 50~\mathrm{mK}$, and $P_L = 1~\mathrm{fW}$, we obtain $G_{\rm QD}^t = P_L/(T_s-T_d) \approx 7\times10^{-15}~\mathrm{W\,K^{-1}}$. The corresponding thermal conductivity is $\kappa_{\rm QD} = G_{\rm QD} l_{\rm QD}/A \approx 2\times10^{-7}~\mathrm{W\,m^{-1}K^{-1}}$. Under the same conditions, the electrical conductance is $\sigma_\text{QD} \approx 0.5~\mu\mathrm{S}$, calculated from the Coulomb peak, yields an effective Lorenz number for the QD, $L = 2G_{\rm QD}/\sigma_\text{QD} (T_s+T_d) \approx 1.1\times10^{-7}~\mathrm{W\,\Omega\,K^{-2}}$, about a factor of four higher than the Wiedemann–Franz value. We note that at the considered average temperature of $T \approx 125~\mathrm{mK}$, the quantum of thermal conductance for a single ballistic channel is $G_q = \pi^2 k_B^2T/3h = 1.2 \times 10^{-13}~\mathrm{W\,K^{-1}}$. Thus, our measured $G_{\rm QD}$ is approximately $6\%$ of this ballistic limit. This significant reduction is consistent with tunnel-limited transport in the Coulomb blockade regime. The suppression of electrical conductance and the observed enhancement of the Lorenz number suggest that the QD acts as an energy filter, selectively allowing high-energy carriers to contribute to heat transport~\cite{Dutta2017}.

\subsection{Nanowire temperature profile}

\begin{figure}
\centering
\includegraphics[width=0.47\textwidth]{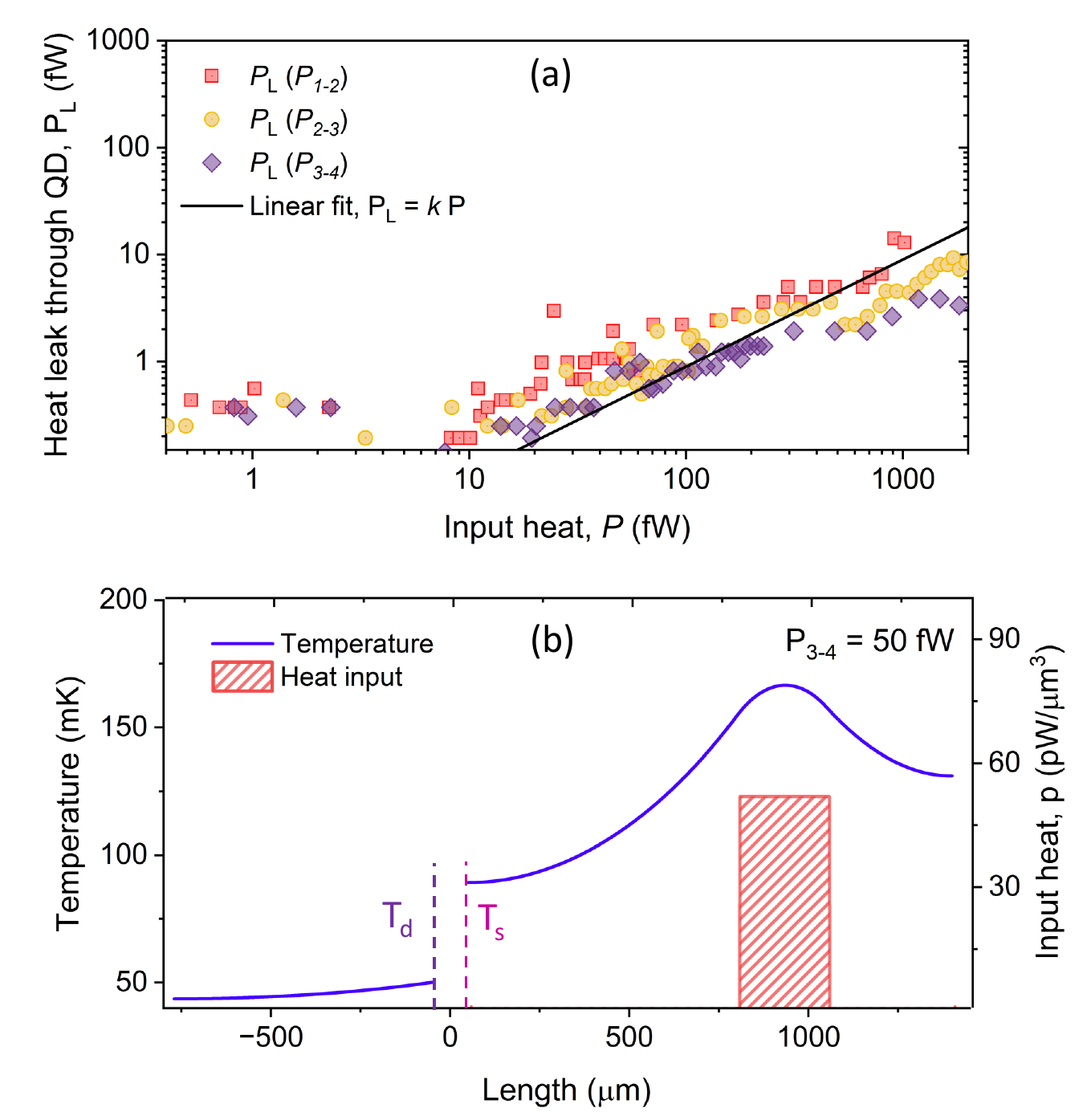}
\caption{(a) The amount of heat flowing through the QD as a function of the input heat to the NW, calculated from the measured drain temperature $T_d$. (b) Calculated temperature profile along the full length of the NW for an applied heating power of $P_H$ = 50 fW. The red bar indicates the heating power density $p$ injected into the NW with leads 3 and 4.}\label{fig4}
\end{figure}

Combining all extracted parameters, we compute the full steady-state temperature profile along the NW for a representative heating power of $50~\mathrm{fW}$ applied between contacts 3 and 4 [Fig.~\ref{fig4}(b)]. The calculation highlights the long-range nature of electronic heat transport and the central role of the heat relaxation length $l_{\rm eq}$ in determining temperature gradients.

\section{Conclusions}

In summary, our measurements establish a quantitative description of heat flow in an InAs nanowire in the clean one-dimensional limit. The calculated electron–phonon exponent $n = 2.6\pm0.2$ indicates energy relaxation with clean one-dimensional coupling. The comparison of electronic heat conduction and phonon cooling shows a characteristic equilibration length $\ell_{\rm eq} \approx 370~\mathrm{nm}$ at $100~\mathrm{mK}$, which sets the length scale above which phonon-mediated thermal transport dominates over heat conduction. Measurement of the temperatures on both sides of the QD further allowed us to determine that less than 1\% of the heat flows through the QD thermometer and that the thermometer exhibits a thermal conductance of only $\sim 6\%$ of the thermal conductance quantum $G_q$. This high degree of thermal isolation ensures minimally invasive local thermometry by quantum dots. These results offer quantitative design rules for thermal management in nanowire-based quantum devices and demonstrate that InAs NW-QD can serve as sensitive, low-loss thermal links for hybrid quantum devices.

\begin{acknowledgments}
We acknowledge fruitful discussions with Claes Thelander, Peter Samuelsson, Martin Leijnse and Simon Wozny and financial support from the Knut and Alice Wallenberg Foundation through the Wallenberg Center for Quantum Technology (WACQT), the European Union (ERC, QPHOTON, 101087343), NanoLund, and the Swedish Research Council (Dnr 2019-04111). Views and opinions expressed are those of the author(s) only and do not necessarily reflect those of the European Union or the European Research Council Executive Agency. Neither the European Union nor the granting authority can be held responsible for them.
\end{acknowledgments}

\bibliography{master}

@article{Stepanov2023,
  title = {Electron-phonon relaxation in a model of a granular film},
  author = {Stepanov, Nikolai A. and Skvortsov, Mikhail A.},
  journal = {Phys. Rev. B},
  volume = {108},
  issue = {20},
  pages = {205415},
  numpages = {9},
  year = {2023},
  month = {Nov},
  publisher = {American Physical Society},
  doi = {10.1103/PhysRevB.108.205415}
}

@article{josefsson2018,
  title={A quantum-dot heat engine operating close to the thermodynamic efficiency limits},
  author={Josefsson, Martin and Svilans, Artis and Burke, Adam M and Hoffmann, Eric A and Fahlvik, Sofia and Thelander, Claes and Leijnse, Martin and Linke, Heiner},
  journal={Nature nanotechnology},
  volume={13},
  number={10},
  pages={920--924},
  year={2018},
  publisher={Nature Publishing Group UK London},
doi={https://doi.org/10.1038/s41565-018-0200-5}
}

@article{mosso2017,
  title={Heat transport through atomic contacts},
  author={Mosso, Nico and Drechsler, Ute and Menges, Fabian and Nirmalraj, Peter and Karg, Siegfried and Riel, Heike and Gotsmann, Bernd},
  journal={Nature nanotechnology},
  volume={12},
  number={5},
  pages={430},
  year={2017},
  publisher={Nature Publishing Group},
    doi={https://doi.org/10.1038/nnano.2016.302}
}

@article{pekola2015,
  title={Towards quantum thermodynamics in electronic circuits},
  author={Pekola, Jukka P},
  journal={Nature physics},
  volume={11},
  number={2},
  pages={118--123},
  year={2015},
  publisher={Nature Publishing Group UK London},
doi={https://doi.org/10.1038/nphys3169}
}

@article{chen2018,
  title={Thermoelectric power factor limit of a 1D nanowire},
  author={Chen, I-Ju and Burke, Adam and Svilans, Artis and Linke, Heiner and Thelander, Claes},
  journal={Physical review letters},
  volume={120},
  number={17},
  pages={177703},
  year={2018},
  publisher={APS},
  doi = {10.1103/PhysRevLett.120.177703}
}

@article{shi2020,
author = {Shi, Xiao-Lei and Zou, Jin and Chen, Zhi-Gang},
title = {Advanced Thermoelectric Design: From Materials and Structures to Devices},
journal = {Chemical Reviews},
volume = {120},
number = {15},
pages = {7399-7515},
year = {2020},
doi = {10.1021/acs.chemrev.0c00026}
}

@article{Giazotto2006,
  title = {Opportunities for mesoscopics in thermometry and refrigeration: Physics and applications},
  author = {Giazotto, Francesco and Heikkila, Tero T. and Luukanen, Arttu and Savin, Alexander M. and Pekola, Jukka P.},
  journal = {Rev. Mod. Phys.},
  volume = {78},
  issue = {1},
  pages = {217--274},
  numpages = {0},
  year = {2006},
  month = {Mar},
  publisher = {American Physical Society},
  doi = {10.1103/RevModPhys.78.217}
}

@article{chen2021,
  title={Non-Fourier phonon heat conduction at the microscale and nanoscale},
  author={Chen, Gang},
  journal={Nature Reviews Physics},
  volume={3},
  number={8},
  pages={555--569},
  year={2021},
  publisher={Nature Publishing Group UK London},
doi={https://doi.org/10.1038/s42254-021-00334-1}
}

@article{Majidi2022,
author = {Majidi, Danial and Josefsson, Martin and Kumar, Mukesh and Leijnse, Martin and Samuelson, Lars and Courtois, Herv{\'e} and Winkelmann, Clemens B. and Maisi, Ville F.},
title = {Quantum Confinement Suppressing Electronic Heat Flow below the Wiedemann–Franz Law},
journal = {Nano Letters},
volume = {22},
number = {2},
pages = {630-635},
year = {2022},
doi = {10.1021/acs.nanolett.1c03437}

}

@article{karimi2021,
    author = {Karimi, Bayan and He, Hans and Chang, Yu-Cheng and Wang, Libin and Pekola, Jukka P. and Yakimova, Rositsa and Shetty, Naveen and Peltonen, Joonas T. and Lara-Avila, Samuel and Kubatkin, Sergey},
    title = {Electron-phonon coupling of epigraphene at millikelvin temperatures measured by quantum transport thermometry},
    journal = {Applied Physics Letters},
    volume = {118},
    number = {10},
    pages = {103102},
    year = {2021},
    month = {03},
    issn = {0003-6951},
    doi = {10.1063/5.0031315}

}

@article{Battisti2024,
    author = {Battisti, Sebastiano and De Simoni, Giorgio and Braggio, Alessandro and Paghi, Alessandro and Sorba, Lucia and Giazotto, Francesco},
    title = {Extremely weak sub-kelvin electron–phonon coupling in InAs on Insulator},
    journal = {Applied Physics Letters},
    volume = {125},
    number = {20},
    pages = {202601},
    year = {2024},
    month = {11},
    issn = {0003-6951},
    doi = {10.1063/5.0225361}
  
}

@article{Fasth2007,
  title = {Direct Measurement of the Spin-Orbit Interaction in a Two-Electron InAs Nanowire Quantum Dot},
  author = {Fasth, C. and Fuhrer, A. and Samuelson, L. and Golovach, Vitaly N. and Loss, Daniel},
  journal = {Phys. Rev. Lett.},
  volume = {98},
  issue = {26},
  pages = {266801},
  numpages = {4},
  year = {2007},
  month = {Jun},
  publisher = {American Physical Society},
  doi = {10.1103/PhysRevLett.98.266801}
}

@article{Svilans2018,
  title = {Thermoelectric Characterization of the Kondo Resonance in Nanowire Quantum Dots},
  author = {Svilans, Artis and Josefsson, Martin and Burke, Adam M. and Fahlvik, Sofia and Thelander, Claes and Linke, Heiner and Leijnse, Martin},
  journal = {Phys. Rev. Lett.},
  volume = {121},
  issue = {20},
  pages = {206801},
  numpages = {6},
  year = {2018},
  month = {Nov},
  publisher = {American Physical Society},
  doi = {10.1103/PhysRevLett.121.206801}
}

@article{Kumar2024,
author = {Kumar, Mukesh and Nowzari, Ali and Persson, Axel R. and Jeppesen, S{\"o}ren and Wacker, Andreas and Bastard, Gerald and Wallenberg, Reine L. and Capasso, Federico and Maisi, Ville F. and Samuelson, Lars},
title = {Hot Carrier Nanowire Transistors at the Ballistic Limit},
journal = {Nano Letters},
volume = {24},
number = {26},
pages = {7948-7952},
year = {2024},
doi = {10.1021/acs.nanolett.4c01197}


}

@article{Wu2013,
author = {Wu, Phillip M. and Gooth, Johannes and Zianni, Xanthippi and Svensson, Sofia Fahlvik and Gluschke, Jan G{\"o}ran and Dick, Kimberly A. and Thelander, Claes and Nielsch, Kornelius and Linke, Heiner},
title = {Large Thermoelectric Power Factor Enhancement Observed in InAs Nanowires},
journal = {Nano Letters},
volume = {13},
number = {9},
pages = {4080-4086},
year = {2013},
doi = {10.1021/nl401501j}
}

@article{Hansen2005,
  title = {Spin relaxation in InAs nanowires studied by tunable weak antilocalization},
  author = {Hansen, A. E. and Bj\"ork, M. T. and Fasth, C. and Thelander, C. and Samuelson, L.},
  journal = {Phys. Rev. B},
  volume = {71},
  issue = {20},
  pages = {205328},
  numpages = {5},
  year = {2005},
  month = {May},
  publisher = {American Physical Society},
  doi = {10.1103/PhysRevB.71.205328}
  
}

@article{hoffmann2009,
  title={Nanoscale thermometry with a quantum dot},
  author={Hoffmann, Eric A and Linke, Heiner},
  journal={Journal of Low Temperature Physics},
  volume={154},
  number={5},
  pages={161--171},
  year={2009},
  publisher={Springer},
doi={https://doi.org/10.1007/s10909-009-9862-6}
}

@article{ahmed2018,
  title={Primary thermometry of a single reservoir using cyclic electron tunneling to a quantum dot},
  author={Ahmed, Imtiaz and Chatterjee, Anasua and Barraud, Sylvain and Morton, John JL and Haigh, James A and Gonzalez-Zalba, M Fernando},
  journal={Communications Physics},
  volume={1},
  number={1},
  pages={66},
  year={2018},
  publisher={Nature Publishing Group UK London},
doi={https://doi.org/10.1038/s42005-018-0066-8}
}

@article{Mickelsen2023,
  title = {Effects of temperature fluctuations on charge noise in quantum dot qubits},
  author = {Mickelsen, D. L. and Carruzzo, Herv\'e M. and Coppersmith, S. N. and Yu, Clare C.},
  journal = {Phys. Rev. B},
  volume = {108},
  issue = {7},
  pages = {075303},
  numpages = {7},
  year = {2023},
  month = {Aug},
  publisher = {American Physical Society},
  doi = {10.1103/PhysRevB.108.075303}

}

@article{Undseth2023,
  title = {Hotter is Easier: Unexpected Temperature Dependence of Spin Qubit Frequencies},
  author = {Undseth, Brennan and Pietx-Casas, Oriol and Raymenants, Eline and Mehmandoost, Mohammad and Madzik, Mateusz T. and Philips, Stephan G. J. and de Snoo, Sander L. and Michalak, David J. and Amitonov, Sergey V. and Tryputen, Larysa and Wuetz, Brian Paquelet and Fezzi, Viviana and Esposti, Davide Degli and Sammak, Amir and Scappucci, Giordano and Vandersypen, Lieven M. K.},
  journal = {Phys. Rev. X},
  volume = {13},
  issue = {4},
  pages = {041015},
  numpages = {18},
  year = {2023},
  month = {Oct},
  publisher = {American Physical Society},
  doi = {10.1103/PhysRevX.13.041015}
 
}

@article{Prete2019,
author = {Prete, Domenic and Erdman, Paolo Andrea and Demontis, Valeria and Zannier, Valentina and Ercolani, Daniele and Sorba, Lucia and Beltram, Fabio and Rossella, Francesco and Taddei, Fabio and Roddaro, Stefano},
title = {Thermoelectric Conversion at 30 K in InAs/InP Nanowire Quantum Dots},
journal = {Nano Letters},
volume = {19},
number = {5},
pages = {3033-3039},
year = {2019},
doi = {10.1021/acs.nanolett.9b00276},

}

@article{Wang2025,
    author = {Wang, Renzong and Xiong, Yucheng and Sun, Jianshi and Zhou, Yongxiang and Song, Guanyao and Chen, Ge and Liu, Xiangjun},
    title = {Remarkably suppressed lattice thermal conductivity of InAs nanowires by surface electron–phonon coupling},
    journal = {Applied Physics Letters},
    volume = {127},
    number = {2},
    pages = {022202},
    year = {2025},
    month = {07},
    issn = {0003-6951},
    doi = {10.1063/5.0277012}
}

@article{Matthews2012,
  title = {Heat flow in InAs/InP heterostructure nanowires},
  author = {Matthews, J. and Hoffmann, E. A. and Weber, C. and Wacker, A. and Linke, H.},
  journal = {Phys. Rev. B},
  volume = {86},
  issue = {17},
  pages = {174302},
  numpages = {8},
  year = {2012},
  month = {Nov},
  publisher = {American Physical Society},
  doi = {10.1103/PhysRevB.86.174302}
}

@article{Chawner2021,
  title = {Nongalvanic Calibration and Operation of a Quantum Dot Thermometer},
  author = {Chawner, J.M.A. and Barraud, S. and Gonzalez-Zalba, M.F. and Holt, S. and Laird, E.A. and Pashkin, Yu. A. and Prance, J.R.},
  journal = {Phys. Rev. Appl.},
  volume = {15},
  issue = {3},
  pages = {034044},
  numpages = {6},
  year = {2021},
  month = {Mar},
  publisher = {American Physical Society},
  doi = {10.1103/PhysRevApplied.15.034044}
 
}

@article{van2024,
  title={Electron Thermometry},
  author={van der Heijden, Joost},
  journal={arXiv preprint arXiv:2403.16305},
  year={2024}
}

@article{Hofmann2020,
  title = {Phonon spectral density in a GaAs/AlGaAs double quantum dot},
  author = {Hofmann, A. and Karlewski, C. and Heimes, A. and Reichl, C. and Wegscheider, W. and Sch\"on, G. and Ensslin, K. and Ihn, T. and Maisi, V. F.},
  journal = {Phys. Rev. Res.},
  volume = {2},
  issue = {3},
  pages = {033230},
  numpages = {6},
  year = {2020},
  month = {Aug},
  publisher = {American Physical Society},
  doi = {10.1103/PhysRevResearch.2.033230}
  
}

@article{Sergeev2000,
  title = {Electron-phonon interaction in disordered conductors: Static and vibrating scattering potentials},
  author = {Sergeev, A. and Mitin, V.},
  journal = {Phys. Rev. B},
  volume = {61},
  issue = {9},
  pages = {6041--6047},
  numpages = {0},
  year = {2000},
  month = {Mar},
  publisher = {American Physical Society},
  doi = {10.1103/PhysRevB.61.6041},
  url = {https://link.aps.org/doi/10.1103/PhysRevB.61.6041}
}

@incollection{altshuler1985,
  title={Electron--electron interaction in disordered conductors},
  author={Altshuler, Boris L and Aronov, A Gh},
  booktitle={Modern Problems in condensed matter sciences},
  volume={10},
  pages={1--153},
  year={1985},
  publisher={Elsevier}
}

@article{Rammer1986,
  title = {Destruction of phase coherence by electron-phonon interactions in disordered conductors},
  author = {Rammer, J\o{}rgen and Schmid, Albert},
  journal = {Phys. Rev. B},
  volume = {34},
  issue = {2},
  pages = {1352--1355},
  numpages = {0},
  year = {1986},
  month = {Jul},
  publisher = {American Physical Society},
  doi = {10.1103/PhysRevB.34.1352},
  url = {https://link.aps.org/doi/10.1103/PhysRevB.34.1352}
}

@article{Ceder2005,
  title = {Nonohmicity and energy relaxation in diffusive two-dimensional metals},
  author = {Ceder, Roy and Agam, Oded and Ovadyahu, Zvi},
  journal = {Phys. Rev. B},
  volume = {72},
  issue = {24},
  pages = {245104},
  numpages = {6},
  year = {2005},
  month = {Dec},
  publisher = {American Physical Society},
  doi = {10.1103/PhysRevB.72.245104},
  url = {https://link.aps.org/doi/10.1103/PhysRevB.72.245104}
}

@article{Peltonen2010,
  title = {Thermal Conductance by the Inverse Proximity Effect in a Superconductor},
  author = {Peltonen, J. T. and Virtanen, P. and Meschke, M. and Koski, J. V. and Heikkil\"a, T. T. and Pekola, J. P.},
  journal = {Phys. Rev. Lett.},
  volume = {105},
  issue = {9},
  pages = {097004},
  numpages = {4},
  year = {2010},
  month = {Aug},
  publisher = {American Physical Society},
  doi = {10.1103/PhysRevLett.105.097004},
  url = {https://link.aps.org/doi/10.1103/PhysRevLett.105.097004}
}

@article{Dutta2017,
  title = {Thermal Conductance of a Single-Electron Transistor},
  author = {Dutta, B. and Peltonen, J. T. and Antonenko, D. S. and Meschke, M. and Skvortsov, M. A. and Kubala, B. and K\"onig, J. and Winkelmann, C. B. and Courtois, H. and Pekola, J. P.},
  journal = {Phys. Rev. Lett.},
  volume = {119},
  issue = {7},
  pages = {077701},
  numpages = {5},
  year = {2017},
  month = {Aug},
  publisher = {American Physical Society},
  doi = {10.1103/PhysRevLett.119.077701},
  url = {https://link.aps.org/doi/10.1103/PhysRevLett.119.077701}
}

@article{Dorsch2021,
doi = {10.1088/1367-2630/ac434c},
url = {https://doi.org/10.1088/1367-2630/ac434c},
year = {2021},
month = {dec},
publisher = {IOP Publishing},
volume = {23},
number = {12},
pages = {125007},
author = {Dorsch, Sven and Fahlvik, Sofia and Burke, Adam},
title = {Characterization of electrostatically defined bottom-heated InAs nanowire quantum dot systems},
journal = {New Journal of Physics}
}

@article{Swinkels2015,
doi = {10.1088/0957-4484/26/38/385401},
url = {https://doi.org/10.1088/0957-4484/26/38/385401},
year = {2015},
month = {sep},
publisher = {IOP Publishing},
volume = {26},
number = {38},
pages = {385401},
author = {Swinkels, M Y and van Delft, M R and Oliveira, D S and Cavalli, A and Zardo, I and van der Heijden, R W and Bakkers, E P A M},
title = {Diameter dependence of the thermal conductivity of InAs nanowires},
journal = {Nanotechnology}
}

@article{Kubala2008,
  title = {Violation of the Wiedemann-Franz Law in a Single-Electron Transistor},
  author = {Kubala, Bj\"orn and K\"onig, J\"urgen and Pekola, Jukka},
  journal = {Phys. Rev. Lett.},
  volume = {100},
  issue = {6},
  pages = {066801},
  numpages = {4},
  year = {2008},
  month = {Feb},
  publisher = {American Physical Society},
  doi = {10.1103/PhysRevLett.100.066801},
  url = {https://link.aps.org/doi/10.1103/PhysRevLett.100.066801}
}
\end{document}